


\def\GeV{\,{\rm GeV}}

\def\cmm2{{\,\rm cm^{-2}}}
\def\cm2{{\,{\rm cm}^2}}
\def\cmm3{{\,{\rm cm}^{-3}}}
\def\gcmm3{{\,{\rm g\,cm^{-3}}}}

\def\mpl{{m_{\rm Pl}}}

\newbox\bigstrutbox
\setbox\bigstrutbox=\hbox{\vrule height12pt depth5pt width0pt}
\def\bigstrut{\relax\ifmmode\copy\bigstrutbox\else\unhcopy\bigstrutbox\fi}

\def\la{\mathrel{\mathpalette\fun <}}

\def\fun#1#2{\lower3.6pt\vbox{\baselineskip0pt\lineskip.9pt
  \ialign{$\mathsurround=0pt#1\hfil##\hfil$\crcr#2\crcr\sim\crcr}}}

\font\tenrm=cmr10 scaled 1200
\font\sevenrm=cmr7 scaled 1200
\font\fiverm=cmr5 scaled 1200
\font\teni=cmmi10 scaled 1200
\font\seveni=cmmi7 scaled 1200
\font\fivei=cmmi5 scaled 1200
\font\tensy=cmsy10 scaled 1200
\font\sevensy=cmsy7 scaled 1200
\font\fivesy=cmsy5 scaled 1200
\font\tenex=cmex10 scaled 1200
\font\tenbf=cmbx10 scaled 1200
\font\sevenbf=cmbx7 scaled 1200
\font\fivebf=cmbx5 scaled 1200
\font\tenit=cmti10 scaled 1200
\font\tensl=cmsl10 scaled 1200
\font\tentt=cmtt10 scaled 1200

\skewchar\teni='177 \skewchar\seveni='177 \skewchar\fivei='177
\skewchar\tensy='60 \skewchar\sevensy='60 \skewchar\fivesy='60

\font\ninei=cmmi10
\font\sixi=cmmi7
\font\fouri=cmmi5
\font\ninesy=cmsy10
\font\sixsy=cmsy7
\font\foursy=cmsy5

\skewchar\ninei='177 \skewchar\sixi='177 \skewchar\fouri='177
\skewchar\ninesy='60 \skewchar\sixsy='60 \skewchar\foursy='60
\def\twelvepoint{\def\rm{\fam0 \tenrm}
  \textfont0=\tenrm \scriptfont0=\sevenrm \scriptscriptfont0=\fiverm
  \rm
  \textfont1=\teni \scriptfont1=\seveni \scriptscriptfont1=\fivei
  \def\mit{\fam1 } \def\oldstyle{\fam1 \teni}
  \textfont2=\tensy \scriptfont2=\sevensy \scriptscriptfont2=\fivesy
  \def\cal{\fam2 }
  \textfont3=\tenex \scriptfont3=\tenex \scriptscriptfont3=\tenex
  \textfont\itfam=\tenit \def\it{\fam\itfam\tenit}
  \textfont\slfam=\tensl \def\sl{\fam\slfam\tensl}
  \textfont\bffam=\tenbf \scriptfont\bffam=\sevenbf
     \scriptscriptfont\bffam=\fivebf \def\bf{\fam\bffam\tenbf}
  \textfont\ttfam=\tentt \def\tt{\fam\ttfam\tentt}
  \baselineskip=14pt}

\outer\def\beginsection#1\par{\bigskip\vbox{\message{#1}\noindent{\bf#1}}
  \nobreak\smallskip\vskip-\parskip\noindent}
\def\exdent#1\par{\noindent\hang\frenchspacing#1\par}
\hsize=6.5in
\vsize=8.8in

\twelvepoint
\baselineskip=16pt
\def\rv{\rho_{\rm vac}}
\def\rp{{r^\prime}}
\def\Op{{\Omega^\prime}}
\def\khat{{\bf{\hat k}}}
\def\xhat{{\bf{\hat x}}}

\rightline{FERMILAB--Pub--92/295-A}
\rightline{October 1992}
\rightline{(submitted to {\it Physical Review D})}
\bigskip
\bigskip
\bigskip

\centerline{\bf Gravitational Radiation from Colliding Vacuum Bubbles:}
\centerline{\bf Envelope Approximation to Many-Bubble Collisions}
\vskip 0.6in
\centerline{Arthur Kosowsky$^{1,3}$ and Michael S.~Turner$^{1,2,3}$}
\bigskip
\bigskip
\centerline{\it Departments of Physics$^1$ and Astronomy \& Astrophysics$^2$}
\centerline{\it Enrico Fermi Institute}
\centerline{\it The University of Chicago}
\centerline{\it Chicago, IL  60637-1433}
\smallskip
\centerline{\it $^3$NASA/Fermilab Astrophysics Center}
\centerline{\it Fermi National Accelerator Laboratory}
\centerline{\it Batavia, IL  60510-0500}

\vskip 0.5in
\noindent{\bf Abstract.} We introduce an approximation to calculate
the gravitational radiation produced by the
collision of true-vacuum bubbles that is simple enough to allow
the simulation of a phase transition by the collision of
hundreds of bubbles.  This ``envelope approximation''
neglects the complicated ``overlap''
regions of colliding bubbles and follows only the evolution
of the bubble walls.
The approximation accurately reproduces previous
results for the gravitational radiation from the
collision of two scalar-field vacuum bubbles.  Using a
bubble nucleation rate given by $\Gamma =
\Gamma_0 e^{\beta t}$, we simulate a phase transition
by colliding 20 to 200 bubbles;
the fraction of vacuum energy released into gravity
waves is $E_{\rm GW}/E_{\rm vac} = 0.06(H/\beta)^2$
and the peak of the spectrum occurs at
$\omega_{\rm max}=1.6\beta$
($H^2=8\pi G\rho /3$ is the Hubble constant associated with
the false-vacuum phase).   The spectrum is very similar
to that in the two-bubble case, except that the efficiency
of gravity-wave generation is about five times higher,
presumably due to the fact that a given bubble collides
with many others.  Finally, we consider two further ``statistical''
approximations, where the gravitational radiation is computed
as an incoherent sum over individual bubbles weighted by
the distribution of bubble sizes.  These approximations
provide reasonable estimates of the gravitational-wave
spectrum with far less computation.

\vfill\eject

\noindent{\bf I. Introduction}
\medskip

The cosmic background of gravitational radiation provides a
unique probe of the early universe. Unlike
electromagnetic radiation, gravity waves propogate virtually
unimpeded since the Planck epoch, providing an unmodified
record of cosmic events. Possible cosmological sources
include the thermal background
(the graviton analogue of the microwave background), inflation [1],
cosmic strings [2], a pregalactic star
population [3], and phase transitions [4]. In particular, strongly
first-order phase transitions are among the most promising of all
these sources [5]: The energy released in gravitational waves can
approach 1\% of that in the ambient thermal bath.

In a recent paper we initiated a detailed investigation of
gravity wave production from strongly
first-order phase transitions by calculating the radiation
from two colliding vacuum bubbles [6].
Beginning with a scalar field configuration
corresponding to two bubbles nucleated simultaneously and far
apart, we used the Klein-Gordon equation to evolve the scalar
field for a time $\tau$ comparable to the initial bubble
separation. (In realistic phase transitions the duration of the
transition is comparable to the typical separation of nucleation
sites [7].) From the scalar-field configuration, we calculated the
stress-energy tensor and, in the linearized gravity approximation,
the energy spectrum of radiated gravity waves.
The pair of vacuum bubbles radiates efficiently: the fraction of
energy that goes into gravity waves is
$${E_{\rm GW}\over E_{\rm vac}}\approx 1.3\times 10^{-3}
\left({\tau\over H^{-1}}\right)^2,\eqno(1)$$
where $E_{\rm vac}$ is the total energy liberated by the
two vacuum bubbles, $\tau$ is the total time of the bubble
evolution, expected to be of order 0.01 to 1 of $H^{-1}$,
and $H^2=8\pi G\rv /3$
is the Hubble parameter associated with the vacuum energy.
The spectrum of radiation peaks at a frequency
$$\omega_{\rm max}\simeq 3.8\tau^{-1}.\eqno(2)$$

These results imply that vacuum bubble collisions can indeed be potent
sources of gravitational radiation. However, our work has its
limitations, the foremost being the
use of a time cutoff in the simulation to model the end of the
phase transition.
Specifically, we smoothly ramped the scalar
field gradients (the source of gravitational radiation) to zero
after a time $\tau$. While such an {\it ad hoc} prescription
greatly simplifies the problem and is probably reasonable, it
clearly neglects multi-bubble effects. The motivation for
colliding many bubbles is to model more realistically
a phase transition.

A direct attack on the many-bubble problem employing
scalar-field evolution is numerically
infeasible. The two-bubble problem was made tractable
by exploiting the $O(2,1)$ symmetry possessed by the space-time
of two vacuum bubbles, which makes the scalar field evolution
effectively one, rather than two, dimensional. Even the slight
generalization to a pair of bubbles nucleated at different
times, a situation still having rotational symmetry about the
axis connecting the two bubbles, proved nearly impossible.
The most general case of many bubbles in three dimensions has no
symmetries and is beyond present computing capabilities.
The problem is difficult numerically because of two
disparate scales. The bubble wall thickness at nucleation is small
compared to the size of the bubble at collision; moreover, the
bubble wall becomes thinner due to Lorentz contraction
as the bubble expands.

To proceed further requires dispensing with the detailed
dynamics of the scalar field. The results of our two-bubble
simulations suggested an elegant approximation.
The spectrum and amount of gravitational radiation
depended only on the gross features of the bubble collisions: the
vacuum energy and the size of the bubbles at the end of the
phase transition ({\it i.e.}, the cutoff time).
Even though the field dynamics after a
bubble collision are quite intricate, the overall contribution
to the radiation from the complicated small-scale motions
adds incoherently and is
subdominant. This prompted us to consider an ``envelope
approximation'': the bubbles are treated as infinitely
thin, and in the regions where bubbles overlap, the bubble wall is
completely ignored. Only the envelope of the evolving bubble network
is considered. As we shall discuss, the envelope approximation very accurately
reproduces our previous results for two colliding bubbles and
allows us to model a phase transition with the collision of
hundreds of bubbles. When applied to a phase transition where
the bubble nucleation rate increases exponentially with time,
$\Gamma\propto\exp{\beta t}$, the fraction of vacuum energy
liberated in gravitational waves is found to be
$${E_{\rm GW}\over E_{\rm vac}}\approx
0.06\left(H\over\beta\right)^2, \eqno(3)$$
or about five times the efficiency estimated from the collision
of two bubbles (for such a nucleation rate, the duration of the
transition is $\tau\sim {\rm few}\,\beta$).

The paper is organized as follows.
The next Section of this paper
describes the envelope approximation, with detailed
comparisons to the previous two-bubble results obtained from scalar-field
evolution [6]. In Section III we first review some pertinent
aspects of bubble nucleation theory [7], and then present our
numerical results for the gravitational radiation from large
numbers of colliding bubbles. In Section IV, we present another
approximation which treats the production of gravitational waves
in a statistical sense: as the incoherent sum of radiation from
individual bubbles, weighted by the
distribution of bubble sizes. Many of the results in Section III can be
reproduced by this simple recipe, and with far less computation.
We finish with a summary of our work and some concluding remarks.
A review of tensor spherical
harmonics and some auxiliary formulas
used in Section IV are relegated to an Appendix.
A summary of the results
of this paper and our earlier work [6], as well as the
application of our results to cosmological phase transitions, is
presented elsewhere [8].

\vfill\eject
\noindent{\bf II. Envelope Approximation}
\nobreak
\medskip
\centerline{\sl (a)  Review of vacuum bubbles}
\nobreak
\medskip

We consider a real scalar field $\varphi$ with a potential possessing two
non-degenerate local minima:
$${\cal L}={1\over 2}\partial^\mu \varphi
\partial_\mu \varphi - V(\varphi).\eqno(4)$$
Throughout we use a metric with signature $(+---)$.  The exact form for
the potential is not important, but where needed we use
$$V(\varphi)={\lambda\over 8}(\varphi^2-\varphi_0^2)^2
+\epsilon\lambda\varphi_0^3(\varphi + \varphi_0).\eqno(5)$$
The dimensionless number $\epsilon$ measures the degree of
symmetry breaking between the two minima near $\pm \varphi_0$. The
relative minimum corresponding to $+\varphi_0$ is the ``false
vacuum,'' while the global minimum corresponding to $-\varphi_0$
is the ``true vacuum.'' The vacuum energy density is defined as
the difference in energy density between the true and false
vacua; here, $\rv\simeq 2\epsilon\lambda\varphi_0^4$. The height
of the potential barrier between the two vacuum states is
$\simeq \lambda\varphi_0^4/8$. The relevant features of the
potential are that it possesses two inequivalent local minima
differing in vacuum energy by $\rv$,
and that the height of the barrier between
the two minima is large enough so that the false vacuum decays
via quantum tunnelling.

Classically, the false-vacuum state is stable, but quantum
effects cause its decay to the true-vacuum state.
This decay proceeds via the quantum nucleation
and expansion of bubbles of the true-vacuum phase which
spontaneously appear from the false-vacuum state. Coleman has
shown that the bubble with minimum action is $O(4)$-invariant in
Euclidean space [9]; the initial bubble profile is
obtained by analytically continuing to Minkowski space and
taking the $t=0$ time slice. The vacuum bubble then evolves
according to the Klein-Gordon equation and has
$O(3,1)$ symmetry; i.e., the scalar field $\varphi$ is a
function only of the quantity $t^2 - x^2 - y^2 - z^2$. The
energy difference between the true and false vacuum phases
creates an effective outward pressure on the bubble wall, causing it
to expand with constant acceleration. For our
purposes, the important aspects of bubble dynamics are that the
expansion speed rapidly approaches the speed of light, and that
the false-vacuum energy liberated becomes kinetic and gradient
energy of the bubble wall [10].

The bubble's symmetry allows us to quantify the above statements
while deriving a result which will be useful in the envelope
approximation.  First, the $O(3,1)$ symmetry immediately implies
that the position of the bubble wall is given by
$${\bf x}_{\rm wall}^2 - t^2 = R_0^2,\eqno(6)$$
where $R_0$ is the initial radius of the bubble and
${\bf x}_{\rm wall}$ denotes a fiducial point within the bubble
wall. Next, consider the stress-energy tensor
associated with the expanding bubble:
$$T_{\mu\nu}({\bf x},t)=\partial_\mu\varphi\partial_\nu\varphi
-g_{\mu\nu}{\cal L}.\eqno(7)$$
The energy density in the scalar
field is given by the time-time component of the stress tensor:
$$T_{00}({\bf x},t)={1\over 2}\Biggl[\left({\partial\varphi\over \partial
t}\right)^2 + \left({\partial\varphi\over \partial r}\right)^2
\Biggr] + V(\varphi)\eqno(8)$$
where we have used the spherical symmetry of the bubble solution.
The bubble's $O(3,1)$ symmetry can be used to write the
the energy of the bubble wall at any time $t$ as
$$\eqalignno{E(t)&\approx
4\pi\int dr\,r^2\Biggl[{1\over 2}\left({\partial\varphi \over
\partial t}\right)^2 + {1\over 2}\left({\partial\varphi\over
\partial r}\right)^2\Biggr]&(9a)\cr
&=2\pi\int_0^\infty ds\left({d\varphi\over ds}\right)^2\Biggl[
{t^2\sqrt{(s^2+t^2)}\over s} + {\sqrt{(s^2+t^2)^3}\over
s}\Biggr],&(9b)\cr}$$
where $s=\sqrt{r^2-t^2}$ and
$\varphi(s)=\varphi(r,t=0)$ is the profile of the initial
bubble solution. Note that for
$r^2<t^2$, $d\varphi/ds=0$ so the integral is zero. The two terms
represent the kinetic and gradient energy of the bubble wall. We
neglect the potential energy term inside the bubble wall as it rapidly
becomes unimportant as the bubble wall gets thinner; we have not
included the false-vacuum energy outside the bubble, as we are
only interested in the energy liberated by the bubble.
In Fig.~1, the kinetic and gradient energy for a bubble are
shown as a function of time. After a short time, the energies
become equal; each is half of $4\pi\rv t^3/3$, the total vacuum
energy liberated by the bubble. From Eq.~(9b) it is simple to
see that gradient and kinetic energies are equal and increase as
$t^3$ for $t\gg R_0$: $d\varphi/ds$ is only nonzero when $s$ is
close to $R_0$; when $t\gg R_0$ each term in the integrand
approaches $(d\varphi/ds)^2 t^3/s$.

\medskip
\centerline{\sl (b) Envelope approximation}
\nobreak
\medskip

As in our previous paper [6], we compute gravity-wave
production in the linearized gravity approximation, valid for
bubble sizes less than $H^{-1}$ (recall $H^2=8\pi G\rv /3$). The
energy radiated in gravitational waves can be expressed in terms
of the Fourier transform of the spatial components of the
scalar-field stress-energy tensor. Further, in computing
$T_{ij} ({\bf k},\omega)$ we may neglect the ${\cal L}g_{ij}$
piece as it is a pure trace and does not act as a source for
gravitational radiation. Thus the fundamental quantity is
$$T_{ij}(\khat,\omega)={1\over 2\pi}\int_0^\infty
dt\,e^{i\omega t}\int d^3 x\partial_i\varphi\partial_j\varphi
e^{-i\omega \khat\cdot{\bf x}}\eqno(10)$$
where $\khat$ is a unit wave-vector. As before,
we adopt Weinberg's unusual normalization convention
for the Fourier transform [11].
The scalar-field configuration of interest is that of many
colliding vacuum bubbles. In the envelope approximation we
assume that the overlap regions where bubbles have expanded
into one another do not contribute
substantially to the gravitational radiation, and exclude these
regions from the spatial integration (see Fig.~2). We can
then break up the
integral into integration regions, one surrounding each
nucleation site and extending out to the bubble radius. Equation
(10) becomes
$$T_{ij}(\khat,\omega)\approx{1\over{2\pi}}\int_0^\infty
dt\,e^{i\omega t}\Biggl[\sum_{n=1}^N\int_{S_n}d\Omega^\prime\,
e^{-i\omega \khat\cdot{\bf x}}\int dr^\prime {r^\prime}^2
\partial_i\varphi\partial_j\varphi \Biggr]\eqno(11)$$
where $N$ is the number of bubbles, $S_n$ is the portion of the
surface of bubble $n$ that remains uncollided at time $t$, and
the primed spherical coordinates are chosen independently around
the center of each bubble. We have also assumed
that the wall thickness is small compared to
$\omega^{-1}$: $\omega\Delta r\ll 1$.  This means $e^{-i\omega
\khat\cdot{\bf x}}$ is essentially constant across the bubble
wall and can be factored out of the $r$-integral. In practice,
for the frequencies of interest, this is always an excellent
approximation.

Next we use the fact that each bubble is spherically symmetric
around its center, so that $\varphi$ is independent of the
angular variables. Dropping the primes for notational convenience,
the stress tensor components become
$$T_{ij}(\khat,\omega)={1\over{2\pi}}\int_0^\infty
dt\,e^{i\omega t}\Biggl[\sum_{n=1}^N e^{-i\omega\khat\cdot{\bf
x}_n}\int_{S_n}d\Omega\,
e^{-i\omega \khat\cdot{\bf x}}\xhat_i\xhat_j\int dr\, r^2
\left(\partial\varphi\over\partial r\right)^2\Biggr]\eqno(12)$$
where ${\bf x}_n$ is the nucleation site of the {\it n}\/th bubble and
$\xhat_i$ is the {\it i}\/th component of a unit vector pointing
from ${\bf x}_n$ in the direction $d\Omega$:
$$\hat x=\sin\theta\cos\phi,\qquad\hat
y=\sin\theta\sin\phi,\qquad\hat z=\cos\theta.$$
Above we showed that the kinetic and gradient energy
associated with a bubble wall
are equal after a small amount of bubble expansion. Using Eq.
(9a), for each bubble we have
$$4\pi\int dr r^2 \left(\partial\varphi\over\partial r\right)^2
\approx{4\pi\over 3}R^3\rv\eqno(13)$$
where the bubble radius $R(t)\approx t$. This is a very good
approximation by the time a bubble has doubled in size; {\it
cf.}\ Fig.~1. Substituting $t^3\rv/3$
for the radial integral in Eq.~(12) leads to
$$T_{ij}(\khat,\omega)={\rv\over{6\pi}}\int_0^\infty
dt\,e^{i\omega t}\Biggl[\sum_{n=1}^N (t-t_n)^3 e^{-i\omega\khat\cdot{\bf
x}_n}\int_{S_n}d\Omega\,
e^{-i\omega \khat\cdot{\bf x}}\xhat_i\xhat_j\Biggr]\eqno(14)$$
where $t_n$ is the nucleation time of bubble $n$.

The total energy radiated in gravity waves is given in terms of
$T_{ij}(\khat, \omega)$ by [11]
$${dE\over d\omega d\Omega}=2G\omega^2\Lambda_{ij,lm}(\khat)
T_{ij}^*({\bf k},\omega) T_{lm}({\bf k},\omega)\eqno(15)$$
where $\Lambda_{ij,lm}$ is the projection tensor for gravity waves,
$$\displaylines{\qquad\qquad
\Lambda_{ij,lm}(\khat)\equiv\delta_{il}\delta_{jm} -
2\khat_j\khat_m\delta_{il} + {\scriptstyle {1\over 2}}\khat_i\khat_j
\khat_l\khat_m \hfill\cr\hfill{}-
{\scriptstyle {1\over 2}}\delta_{ij}\delta_{lm} +
{\scriptstyle {1\over 2}}\delta_{ij}\khat_l\khat_m +
{\scriptstyle {1\over 2}}
\delta_{lm}\khat_i\khat_j.\qquad\qquad(16)\cr}$$

\medskip
\centerline{\sl (c) Scaling properties}
\nobreak
\medskip

{}From Eqs. (14) and (15), two important scaling relations are
evident. First, the total radiated energy is explicitly
proportional to $\rv^2$, just as we found previously
in the two-bubble case [6]. Second,
since the bubbles expand at essentially the speed of light
and to a good approximation have zero
thickness and zero initial size, the problem has no intrinsic
length/time scale.
Making the transformation $t\rightarrow \gamma t$,
${\bf x}\rightarrow \gamma {\bf x}$, we find the following
scaling properties:
$$\omega\rightarrow {\omega\over \gamma};\eqno(17a)$$
$${dE\over d\omega d\Omega}\rightarrow \gamma^6{dE\over d\omega
d\Omega} ;\eqno(17b)$$
$$E_{\rm GW}\rightarrow \gamma^5 E_{\rm GW}.\eqno(17c)$$
The length/time scale is set by the average separation between bubble
nucleation sites.

Equations (17) show that the
total energy radiated from a volume containing a fixed number of
bubbles will vary with the fifth power of the mean bubble
separation. As we will show in the next Section, these scalings can
also be expressed in terms of the bubble nucleation rate, since
the typical separation of nucleation sites is determined by the
nucleation rate.
For the two-bubble case, the above scalings were found previously
to hold to very good accuracy [6],
where the relevant scale is the total evolution time.
These scalings have great practical importance as they allow
us to apply the results of a single numerical simulation to any
phase transition with a bubble nucleation rate of the same
functional form. In particular, we shall use the functional form
$\Gamma\propto e^{\beta t}$---a nucleation rate that increases
exponentially with time---in which case the length/time scale
is just $\beta^{-1}$.

\bigskip
\centerline{\sl (d) Two bubbles, quadrupole approximation}
\nobreak
\medskip

In order to determine the accuracy of the envelope
approximation, we compare the results it gives for two bubbles
to the results calculated previously
with the exact scalar field evolution [6].
As a warm-up, we begin with the conventional
``quadrupole approximation,''
corresponding to the limit ${\bf k\cdot x}\rightarrow 0$:
$$T_{ij}(\khat,\omega)={\rv\over{6\pi}}\int_0^\infty
dt\,e^{i\omega t}\Biggl[\sum_{n=1}^N (t-t_n)^3
\int_{S_n}d\Omega\,\xhat_i\xhat_j\Biggr].\eqno(18)$$
Note that because our source is not small compared to the
wavelength of the radiation, this approximation does {\it not}
correspond to the quadrupole term in the multipole expansion;
see Ref. [6] for detailed elucidation of this point.
The quadrupole approximation simplifies the requires
calculations, and in the limit $\omega\rightarrow 0$ gives the
correct result; thus we use it as a starting point for our
comparisons.

Consider two bubbles of negligible initial size
nucleated simultaneously at $t=0$ on the $z$-axis at $\pm
d/2$. The bubbles will first ``kiss'' at $t=d/2$. Define
$\cos\alpha =d/2t$ for $t>d/2$ and $\alpha=0$ for $t<d/2$;
$\alpha$ is the angle
excluded from the angular integration because of bubble overlap
in the envelope approximation. Then
$$T_{ij}={\rv\over 6}\int_0^\infty dt e^{i\omega t} t^3\Biggl[
\int_0^{\pi-\alpha}d\theta\sin\theta\int_0^{2\pi}d\phi\xhat_i\xhat_j
+ \int_\alpha^\pi d\theta\sin\theta\int_0^{2\pi}d\phi\xhat_i\xhat_j
\Biggr].\eqno(19)$$
Using the spherical coordinates defined in Eq.~(11),
$$T_{xx}=T_{yy}={\rv\over 3}\int_{d/2}^\infty dt\,e^{i\omega t}
\left({2\over 3}t^3 + {d\over 2} t^2 -{d^3\over 24}\right)C(t)\eqno(20a)$$
and
$$T_{zz}={2\rv\over 9}\int_{d/2}^\infty dt\,e^{i\omega t}
\left(t^3 + {d^3\over 8}\right)C(t),\eqno(20b)$$
where we have included a time cutoff function $C(t)$
which decreases smoothly from 1 to 0 on a time scale
$\tau\sim{\cal O}(d)$. As mentioned earlier, the
cutoff was introduced in our previous two-bubble calculations to
model the completion of the phase transition; $C(t)$ is
discussed in detail in Ref. [6].

Rotational symmetry around the $z$-axis implies the
off-diagonal components of the stress tensor are zero, and
$T_{ij}$ must be of the form
$$T_{ij}=D(\omega)\delta_{ij} + \Delta(\omega)
\delta_{iz}\delta_{jz}.\eqno(21)$$
The first term, being a pure trace, does not contribute to
gravitational radiation. The second term is
$$\eqalignno{\Delta(\omega)&\equiv T_{zz}-{1\over
2}(T_{xx}+T_{yy})\cr &={\rv\over 3}\int_{d/2}^\infty dt
e^{i\omega t}\left({d^3\over 8}-{d\over
2}t^2\right)C(t).&(22)}$$
Substitution into Eq.~(15) gives
$${dE\over d\omega
d\Omega}=G\omega^2|\Delta(\omega)|^2\sin^4\theta;\qquad
{dE\over d\omega}={32\pi\over
15}G\omega^2|\Delta(\omega)|^2.\eqno(23)$$

The comparison of the envelope approximation with the previous
calculations using scalar-field evolution is shown in Fig.
3. (For reference, in the scalar-field evolution case we used
$\tau/d =1.2$ and a gaussian roll-off in the final 10\% of the
evolution time for $C(t)$; see [6].)
The features of the spectrum are reproduced remarkably well. The
overall normalization of the envelope approximation is high by
20 percent. It may seem strange that the total power
radiated is higher from an approximation which neglects a
chunk of the source. However, when two bubbles collide, a reflected
wave begins to propogate outward from the point of collision.
This wave takes the approximate shape that the colliding portion
of the bubbles would have had they not collided. By
neglecting the interaction region, we actually make a given
bubble {\it less} spherical, and hence increase the amount of
radiation.

\bigskip
\centerline{\sl (e) Two bubbles, full linearized gravity
approximation}
\nobreak
\medskip

For two bubbles in the full linearized-gravity approximation, we can
derive formulas analogous to those used with the scalar field
evolution. As in the quadrupole case, let the bubbles be
nucleated at $t=0$ and at $z=\pm d/2$, with $\tau/d=1.2$ and the
same time cutoff function. The problem possesses
rotational symmetry about the $z$-axis, so without loss of
generality we take $k_y=0$, $k_x=\omega\sin\xi$,
$k_z=\omega\cos\xi$. Using the same conventions as in the
quadrupole case, the stress-energy tensor components are given by
$$\displaylines{\qquad T_{ij}(\khat,\omega)=
{\rv\over 6\pi}\int_0^\infty dt e^{i\omega t}
t^3 C(t)\Biggl[ e^{-ik_zd/2}
\int_0^{\pi-\alpha}d\theta\sin\theta\int_0^{2\pi}d\phi
e^{-i{\bf k}\cdot{\bf x}}\xhat_i\xhat_j\hfill\cr
\hfill{} + e^{ik_zd/2}
\int_\alpha^\pi d\theta\sin\theta\int_0^{2\pi}d\phi
e^{-i{\bf k}\cdot{\bf x}}\xhat_i\xhat_j\Biggr].\qquad(24)\cr}$$
The $\phi$ integral can be done explicitly using the identity
$$\int_{-\pi}^{\pi}e^{i\beta\cos x}\cos nx\,dx = 2i^n\pi
J_n(\beta),\eqno(25)$$
resulting in the following expressions:
$$\displaylines{\qquad T_{xx}(\khat,\omega)={\rv\over 3}\int_0^\infty
dt\,e^{i\omega t} t^3 C(t)\int_0^{\pi -\alpha}d\theta\,
\sin^3\theta\cos(k_zt\cos\theta+{\scriptstyle {1\over 2}}k_zd)
\hfill\cr\hfill{}\times
\left[J_0(k_xt\sin\theta)-J_2(k_xt\sin\theta)\right],\qquad (26a)\cr}$$
$$\displaylines{\qquad T_{yy}(\khat,\omega)={\rv\over 3}\int_0^\infty
dt\,e^{i\omega t} t^3 C(t)\int_0^{\pi -\alpha}d\theta\,
\sin^3\theta\cos(k_zt\cos\theta+{\scriptstyle {1\over 2}}k_zd)
\hfill\cr\hfill{}\times
\left[J_0(k_xt\sin\theta)+J_2(k_xt\sin\theta)\right],\qquad (26b)\cr}$$
$$\displaylines{\qquad T_{zz}(\khat,\omega)={2\rv\over 3}\int_0^\infty
dt\,e^{i\omega t} t^3 C(t)\int_0^{\pi -\alpha}d\theta\,
\sin\theta\cos^2\theta\hfill\cr\hfill{}
\times\cos(k_zt\cos\theta+{\scriptstyle {1\over 2}}k_zd)
J_0(k_xt\sin\theta),\qquad (26c)\cr}$$
$$\displaylines{\qquad T_{xz}(\khat,\omega)={-2\rv\over 3}\int_0^\infty
dt\,e^{i\omega t} t^3 C(t)\int_0^{\pi -\alpha}d\theta\,
\sin^2\theta\cos\theta\hfill\cr\hfill{}
\times\sin(k_zt\cos\theta+{\scriptstyle {1\over 2}}k_zd)
J_1(k_xt\sin\theta).\qquad (26d)\cr}$$
Note that $T_{xy}=T_{yz}=0$. The energy radiated in gravity
waves simplifies to
$$\displaylines{\qquad{dE\over d\omega d\Omega}=G\omega^2\Bigl|
T_{zz}(\khat,\omega)\sin^2\xi +T_{xx}(\khat,\omega)\cos^2\xi
\hfill\cr \hfill{}-T_{yy}(\khat,\omega)-2T_{xz}(\khat,\omega)
\sin\xi\cos\xi\Bigr|^2.\qquad (27)\cr}$$
In Fig.~4 we compare the envelope approximation with the
previous results using scalar-field evolution. As
in the quadrupole case, the agreement is excellent, with the
envelope approximation power being slightly greater.

\bigskip
\noindent{\bf III. Numerical Methods and Results}
\nobreak
\medskip
\centerline{\sl (a) Bubble nucleation [7]}
\nobreak
\medskip

The envelope approximation closely reproduces the gravitational
radiation from a two-bubble collision, even showing the same
features in the spectrum. This gives us confidence to apply the
approximation to the situation where many bubbles are nucleated,
collide, and transform all of space to the true vacuum.
In a phase transition the bubble nucleation rate per unit volume,
$\Gamma$, is in general a function of time, due to its dependence
upon the temperature of the universe or the evolution of other
fields. Since very generally $\Gamma$ is the exponential of
some action, we write it as
$$\Gamma(t)=Ce^{-A(t)}\,.\eqno(28)$$
The prefactor $C$ is expected to be
of the order of ${\cal M}^4$, where $\cal
M$ is the mass or energy scale characterizing the transition.
The tunnelling action $A(t)$ must be greater than order unity;
otherwise, the transition will not proceed via bubble nucleation
but through spinodal decomposition, because of the very small
potential barrier between the false and true vacuum states.
(For the form of the potential in Eq.~(5), $A\gg 1$ obtains for
$\epsilon \ll 1$.)
The fact that the nucleation rate varies with time is crucial to
the completion of the phase transition; moreover, how fast it
varies with time determines the distribution of bubble sizes. As
a rough rule, the phase transition completes when one
bubble is nucleated per Hubble volume per Hubble time, {\it
i.e.} when $\Gamma(t)/H^4\sim 1$.  Denote the completion time, about which
we shall be more specific, by $t_*$.  Expanding $A(t)$ about
$t_*$ gives
$$A(t)\approx A_* -\beta(t-t_*),\qquad
\beta = -\left({d\ln\Gamma\over dt}\right)\Biggl|_{t_*}
=\left({\partial A\over \partial t}\right)\Biggr|_{t_*},\eqno(29)$$
where $A_*\equiv A(t_*)$.
In any sensible model, $\beta>0$; {\it i.e.}, the nucleation
rate grows with time. As we shall see, $\beta^{-1}$ sets the
time/length scale for the phase transition.

We now derive some important results for the exponential
nucleation rate. The fundamental quantity is $p(t)$, the
probability that a given point in space remains in the false
vacuum at time $t$. It is given by $p(t)=\exp[-I(t)]$, where
$I(t)$ is the expected fraction of space occupied by
true-vacuum bubbles at time $t$, without regard to
bubble overlap:
$$I(t)={4\pi\over 3}\int_{t_0}^t dt^\prime\Gamma(t^\prime)
a^3(t^\prime)r^3(t,t^\prime),\eqno(30a)$$
where $a(t)$ is the cosmic scale factor, $r(t,t^\prime)$ is
the coordinate radius at time $t$ of a bubble nucleated at time
$t^\prime$, and $t_0$ is the time at which the phase transition
begins [12]. For simplicity and consistency
with our previous neglect
of the expansion of the universe, we take $a$ to be constant and
$ar(t,t^\prime)=t-t^\prime.$ The neglect of the expansion is
justified provided the duration of the transition
is less than $H^{-1}$.
The second assumption implies that bubbles expand to a size far
greater than that when nucleated, which is well-justified in the
cases of interest. Taking $t_0\rightarrow -\infty$ with little
error, it follows that
$$I(t)\approx {8\pi\over \beta^4}\Gamma(t).\eqno(30b)$$
Then the false-vacuum fraction is given by
$$p(t)=e^{-I(t)}\approx e^{-8\pi\Gamma(t)/\beta^4}\eqno(31)$$
where the exponentiation accounts for the bubble overlap. From
$p(t)$ we can compute the duration of the phase transition
and distribution of bubble sizes. The start
and end of the transition are somewhat difficult to define
precisely, but this ambiguity is not important. To be specific,
we can define the start of the transition to be the time $t_m$
when $p(t_m)=e^{-m}\approx 1$, {\it i.e.}, $m\ll 1$. Similarly,
we define the end of the transition to be the time $t_*$ when
$p(t_*) =e^{-M}\approx 0$, {\it i.e.} $M\gg 1$. The duration of
the transition is thus
$$\delta t\equiv t_* - t_m =\ln\left({M\over
m}\right)\beta^{-1},\eqno(32)$$
and depends only logarithmically upon the precise definition of
the start and end of the phase transition. Note that the
duration of the phase transition is set by $\beta^{-1}$, and
that for consistency our neglect of the expansion of the
universe requires the duration to be less than a Hubble time:
$\beta^{-1} \ll H^{-1}$.

The density (per unit volume)
of bubbles of a given radius $r$ at time $t$ is related to
$\Gamma(t)$ and $p(t)$ by
$$\eqalignno{\left({dn\over dr}\right)_t &=p(t^\prime)
\Gamma(t^\prime)\biggl|_{t^\prime=t-r}\cr
&\approx{\beta^4 I(t)\over 8\pi}\exp\bigl[-I(t)e^{-\beta r}
-\beta r\bigr].&(33)}$$
The distribution of bubble sizes attains its maximum at
$${\bar r}(t)={1\over \beta}\ln I(t)\eqno(34)$$
and has a width of order $\beta^{-1}$. In discussing
gravitational wave production it is more
appropriate to examine the energy-weighted bubble distribution.
Since the energy carried in the expanding wall of a bubble is
proportional to its volume, this distribution is obtained by
multiplying $dn/dr$ by $r^3$:
$$\left({dn_E\over dr}\right)_t ={\beta^4 I(t)\rv\over 6}r^3
\exp\left[-I(t)e^{-\beta r} -\beta r\right].\eqno(35)$$
This distribution is peaked at a radius twice as large as
$dn/dr$ (see Fig.~5).

Finally, how is the key parameter $\beta$ related to $H$ and
$\cal M$? Since $H^{-1}$ sets the scale for all time evolution
in the universe, on very general grounds we expect
$\beta=-(\partial A/\partial t)_{t_*}$ to be of the order of
$A(t_*)/H^{-1}$, or $\beta^{-1}\sim H^{-1}/A_*$. If the
transition is to proceed via vacuum bubbles, $A_*$ must be much
greater than one, so the assumption that the transition is
``fast'', $\beta^{-1}\ll H^{-1}$, should generally be satisfied.
We can also estimate $A_*$: $\Gamma(t_*)={\cal M}^4 e^{-A_*} \sim
H^4 \sim {\cal M}^8/\mpl^4$, which implies that $A_*$ should be
of order $\ln(\mpl/{\cal M})$.

\medskip
\centerline{\sl (b) Numerical results}
\nobreak
\medskip

In the previous sub-section we have motivated the use of an
exponential nucleation rate; specifically,
$$\Gamma(t)=\Gamma_0 e^{\beta t}.\eqno(36)$$
As we have discussed, $\beta^{-1}$ sets the fundamental time/length
scale: both the duration of the transition and the typical
bubble size are of order a few $\beta^{-1}$. We use a spherical
volume, and choose $\Gamma_0$ so that on
average, the desired number of bubbles are nucleated in the
sample volume. Our main calculations (see below) use
a sample volume with a radius of $4.46 \beta^{-1}$ and
$\Gamma_0=1.38\times 10^{-3} \beta^4$.
These parameters yield an average
of around thirty bubbles in the sample volume.
For our five different nucleation runs, the
time at which the phase transition completes varied from
$5\beta^{-1}$ to $6\beta^{-1}$, with an average of
$5.63\beta^{-1}$. (We use the time when the last bubble is
nucleated in the sample volume as the completion time
of the phase transition.)

A ``nucleation run'' proceeds as follows. For each time step we: (1)
Perform a Monte Carlo integration over the sample volume to
determine the fraction still in the false-vacuum state. This
takes into account all true-vacuum bubbles nucleated prior to
the time step. (2) Multiply $\Gamma(t)$ by the sample volume
still in the false vacuum and length of the
time step to get the mean number of bubbles nucleated during the
time step. (3) Use this mean number to generate a
Poisson-distributed random number of bubbles nucleated during
the time step. (4) Nucleate this number of bubbles at random
points in
the false-vacuum and during the time step. We choose the time step so
that the time from the first nucleation to the disappearance of all
the false vacuum is around 100 time steps.

The initial conditions for the simulations consist of
nucleation sites and times within the spherical volume produced
by the above procedure. We use a spherical boundary
to minimize edge effects.
We also utilize reflecting boundaries: When a bubble
expands into the boundary, the part of the bubble that passes
the boundary is ignored. In the context of the envelope approximation,
this is equivalent to having another
bubble outside the boundary expand into the bubble in question,
with their first contact occurring at the boundary surface.
Note that in the envelope approximation, reflecting and absorbing
boundaries are equivalent.

The aim of the simulation is to numerically evaluate Eq.~(14)
for the stress-tensor components $T_{ij}(\khat,\omega)$ for the
given initial conditions, and then Eq.~(15) for the spectrum of
gravitational radiation produced. Note no radiation is produced
before the first bubble collision occurs or after the
transition to true vacuum has completed; thus these are
the limits for the time integration. Our numerical calculations
proceed as follows: First we choose values for
$\omega$ and the direction $\khat$. Then we numerically
evaluate the integrals in Eq.~(14). The time partition is chosen for
around 100 time steps, and the angular partition is chosen to give
20 divisions per wavelength at a given radiation frequency.
These partitions produce numerical
integration results accurate to within 5\% at all frequencies of
interest.

Since bubble nucleation is an inherently random process, we want
to find the power spectrum of gravity waves, averaged over many
different nucleation realizations. We have calculated the
radiation from five simulations averaging 30
bubbles each, ranging from 17 to 38 bubbles. Figure 6 shows a
cross-sectional slice through the equator of the spherical
volume for one nucleation run; the combined bubble envelope, which is the
radiation source, is shown at several times.
For each simulation, we have
computed the radiation in six directions, along the $\pm x$,
$\pm y$, and $\pm z$ axes. The results
for $dE/d\omega d\Omega$ for several representative directions are
displayed in Fig.~7.

Several features are evident from these results. First, all of
the spectra peak at a characteristic frequency which roughly
corresponds to the size of the largest bubbles at the end of the
transition. Second, the power generally increases as the number
of bubbles in a given volume decreases, as we expect
($dE/d\omega d\Omega$ should vary as $N^{-5/3}$). Finally and most
obviously, the various runs and directions of observations have
very large fluctuations in both total power and spectrum shape.
This is because with the relatively small
number of bubbles, the source
has large inhomogeneities on the scale of the sample volume.
One direction in one simulation ({\it i.e.}, one out of 30
probes of $dE/d\omega d\Omega$) exhibits a
``beaming'' effect, with ten to a hundred times more power than
average at frequencies higher than several times the
peak frequency.

Figure 8 shows the power radiated in gravity waves per octave
averaged over all five simulations, six directions per
simulation (excluding the one anomolously ``hot'' direction);
the error bars indicate the
statistical deviation of the mean. On the same plot we show our
previous results for the collision of two scalar-field bubbles
[6]. Recall the previous results depend upon the cutoff time
$\tau$, which corresponds to the duration of the transition. To
compare the old results with the present ones, we set
$\tau=3.2/\beta$. We choose this value for $\tau$ because it is
the mean energy-weighted bubble size at the end of the phase
transition ({\it cf.}\ Fig.~5). We have
divided our results by the total vacuum energy of the sample
volume, so
the results shown are the fraction of false-vacuum energy
liberated in gravitational waves per octave.
The two-bubble and many-bubble results are remarkably similar:
behavior at high and low frequencies is almost
identical, and the peaks are at almost the same
frequency. The overall normalization of the many-bubble case is
higher, by a factor of five or so on the low-frequency side and
by an order of magnitude on the high-frequency side.
This increase is not unexpected: In the many-bubble
case, each bubble collides with many others, increasing its
total radiation. The excess of high frequency power also makes
sense, since the size of a given bubble roughly determines the
frequency at which it radiates. In the two-bubble case, both
bubbles are the same size; in the many-bubble case,
smaller bubbles are nucleated late and increase the high-frequency
power.

The total fraction of vacuum energy released in gravitational
radiation is computed by
integrating $dE/d\omega$ and dividing by the
total vacuum energy released:
$${E_{\rm GW}\over E_{\rm vac}}=0.50{G\rv\over\beta^2}
=6.0\times 10^{-2}\left({H\over\beta}\right)^2.\eqno(37)$$
The peak of the energy spectrum is given by $\omega_{\rm
max}\approx 1.6\beta$. We note that the characteristic frequency
defined by the maximum of the energy-weighted bubble
distribution $\omega_*=2\pi/{\bar r}_E\approx 1.6\beta$
coincides with $\omega_{\rm max}$. Comparing these results with
our previous two-bubble results (taking $\tau=3.2/\beta$), the
peak of the spectrum occurs at the same frequency, while the
total fraction of energy liberated in gravity waves is about a
factor of five larger.

To verify that our results are not dominated by edge effects, we
ran one large calculation in a spherical sample volume with
twice the radius of the above simulation and the same nucleation
rate, 180 bubbles in all. The radiation spectrum $dE/d\omega$,
integrated over the six observation directions, is plotted in Fig.~8.
The shape of the spectrum is close to
the average of the previous cases, and
$E_{GW}/E_{\rm vac}=0.47 G\rv/\beta^2$. The close correspondence
(only 5\% difference) with our smaller simulations
demonstrates that edge effects are not important.

\bigskip
\noindent{\bf IV. Statistical Bubble Approximation}
\nobreak
\medskip

Now we discuss two approximations for computing
the gravitational-wave
production from the collision of many bubbles as the incoherent
sum of the radiation from individual bubbles. To this end, we
use the multipole-radiation formalism and the envelope
approximation applied to a typical bubble of size $R$, and
integrate over the distribution of bubble sizes.
The advantage of such an approach is ease of
computation: We can immediately calculate the radiation spectrum
for a given nucleation rate, without recourse to a many-bubble
simulation. Of course, this approach neglects coherent
effects between bubbles and is less accurate than our
calculations of the previous Section,
but as we shall see it still gives a reasonable estimate
of the gravitational radiation produced.

Consider a single expanding bubble in a sea of bubbles.
In the envelope approximation, when the bubble in question
begins interacting with other bubbles, portions
of the bubble surface are ``eaten.'' Neglecting the interaction
regions between bubbles, the stress tensor for what remains of
the bubble is
$$T_{ij}({\bf x},t)=\xhat_i\xhat_j\left({\partial\varphi(r,t)
\over\partial r}\right)^2\Theta(\Omega,t)\eqno(38)$$
where $\Theta$ is defined as
$$\Theta(\Omega,t)=\cases{1,&surface in direction $\Omega$
remains uncollided\cr
0,&otherwise.\cr}\eqno(39)$$
When the bubble wall has completely disappeared
due to collisions with other bubbles, the function $\Theta$ is
zero. This means that an individual bubble can be treated as a
source that is bounded in space and time: $\Delta t,\,\Delta x
\la {\cal O}(\tau)$, where $\tau$ is the duration of the phase transition.
The size and energy density of a given bubble at any time during
a phase transition is known; the function $\Theta$
varies from bubble to bubble. Our statistical approximation boils
down to estimating an ``average'' $\Theta(\Omega,t)$.

\medskip
\centerline{\sl (a) Multipole-radiation formalism}
\nobreak
\medskip

Using the tensor spherical harmonics
presented in the Appendix, we can expand the
transverse-traceless part of the metric perturbation
({\it i.e.}, the gravity wave piece) in the far
field zone ($r\gg \tau$) as follows [13]:
$$h_{ij}^{TT}(t,{\bf r})={G\over r}\sum_{l=2}^\infty \sum_{m=-l}^l
\left[{d^l\over dt^l}I^{lm}(t-r){\rm T}^{E2,lm}_{ij}(\Omega)
+ {d^l\over dt^l}S^{lm}(t-r){\rm
T}^{B2,lm}_{ij}(\Omega)\right],\eqno(40)$$
where the $I^{lm}$ and the $S^{lm}$ are the ``mass'' and
``current'' multipole moments of the source, respectively,
defined by
$$\displaylines{\qquad{d^l\over dt^l}I^{lm}(t)=8(-i)^{l+2}
\int \rp^2 d\rp d\Op dt^\prime d\omega\,
e^{-i\omega(t-t^\prime)}\tau_{pq}(t^\prime,\rp,\Op)\hfill\cr
\hfill \times\biggl[
a_{-2}(l){\rm T}^{2\,l-2,lm}_{pq}(\Op)^* j_{l-2}(\omega \rp)
-a_0(l){\rm T}^{2\,l,lm}_{pq}(\Op)^* j_l(\omega \rp)\hfill\cr
\hfill +a_2(l){\rm T}^{2\,l+2,lm}_{pq}(\Op)^*j_{l+2}(\omega \rp)
\biggr],\qquad(41a)\cr}$$
$$\displaylines{\qquad{d^l\over dt^l}S^{lm}(t)=8(-i)^{l+2}
\int \rp^2 d\rp d\Op dt^\prime d\omega\,
e^{-i\omega(t-t^\prime)}\tau_{pq}(t^\prime,\rp,\Op)\hfill\cr
\hfill\times\biggl[
a_{-1}(l){\rm T}^{2\,l-1,lm}_{pq}(\Op)^*j_{l-1}(\omega \rp)
-a_1(l){\rm T}^{2\,l+1,lm}_{pq}(\Op)^*j_{l+1}(\omega \rp)
\biggr].\qquad(41b)\cr}$$
The tensors ${\rm T}^{2\,i,lm}(\Omega)$ and the coefficients $a_i$ are
given by Equations (A.2) and (A.4) in the Appendix,
$j_l$ is the spherical Bessel function of order $l$, $p$
and $q$ are indices of the tensor components, and $\tau_{pq}$ is
the sum of the stress-energy tensor of matter ($T_{pq}$) and the
Landau-Lifshitz pseudotensor for the effective stress-energy of
the gravitational field. Since we are using the linearized
gravity (appropriate since all gravitational effects are weak;
see Ref. [6]), $\tau_{pq} =T_{pq}$.
The effective stress-energy tensor for gravity waves is
$$T_{\alpha\beta}^{\rm GW} = {1\over 32\pi G}\sum_{i,j}
\langle \partial_\alpha h^{TT}_{ij}\partial_\beta h^{TT}_{ij}
\rangle,\eqno(42)$$
and the power radiated in terms of the multipoles:
$$P(t)={G\over 32 \pi} \sum_{l,m}\biggl\langle
\left|{d^{l+1}\over dt^{l+1}}I^{lm}\right|^2
+\left|{d^{l+1}\over dt^{l+1}}S^{lm}\right|^2
\biggr\rangle,\eqno(43)$$
where $\langle\,\cdot\,\cdot\,\cdot\,\rangle$ indicates a spatial
average over several wavelengths.

The expressions (41) can be simplified considerably. First, only
the exponential factor and the Bessel functions depend on
$\omega$. The $\omega$ integral can be performed explicitly
using the identities
$$\int_{-\infty}^\infty dx\,e^{-ixy}j_n(x)=\cases{
(-i)^n\pi P_n(y),&$|y|<1$;\cr
0,&$|y|>1$,\cr}\eqno(44)$$
where $P_n$ are the Legendre polynomials. This gives
$$\eqalignno{{d^l\over dt^l}I^{lm}(t)&=8\pi(-1)^l
\int_0^\infty \rp^2 d\rp \int d\Op \int_{-1}^1 d\eta\,
\tau_{pq}(t-\eta \rp,\rp,\Op)\biggl[
a_{-2}(l){\rm T}^{2\,l-2,lm}_{pq}(\Op)^* P_{l-2}(\eta)\cr&\qquad
+a_0(l){\rm T}^{2\,l,lm}_{pq}(\Op)^* P_l(\eta)
+a_2(l){\rm T}^{2\,l+2,lm}_{pq}(\Op)^*P_{l+2}(\eta)\biggr].&(45)}$$
Now for a single bubble nucleated at $t=0$,
the $O(3,1)$ symmetry of the scalar-field bubble
configuration allows transformation of the radial
integral:
$$\eqalignno{{d^l\over dt^l}I^{lm}(t)
&=8\pi(-1)^l\int_{-1}^1 d\eta\int_0^\infty ds
\left(\sqrt{t^2+s^2(1-\eta^2)}-t\eta\right)^2
{\sqrt{t^2+s^2(1-\eta^2)}\over s(1-\eta^2)^2}
\left({d\varphi\over ds}\right)^2\cr&\qquad\times
\left[\sum_{i=-2,0,2} P_i(\eta)\int d\Op A^{l+i,lm}(\Op)
\Theta\left(\Op,{t-\eta\sqrt{t^2+s^2(1-\eta^2)}\over
1-\eta^2}\right) \right],&(46)}$$
where
$$A^{l^\prime,lm}(\Op)\equiv a_{l^\prime-l}(l)
{\rm T}^{2\,l^\prime,lm}_{pq}(\Op)^*\xhat_p(\Op)\xhat_q(\Op).
\eqno(47)$$
Here $s=\sqrt{r^2-(t-r\eta)^2}$ is the only quantity the scalar field
depends upon, $\phi^\prime(s)=\partial\phi/\partial r(r,t=0)$,
and $\xhat$ is the unit vector in the direction of $\Op$.
Explicit expressions for the $A^{l^\prime,lm}$ are given in the Appendix.
The further substitution
$$y={t-\eta\sqrt{t^2+s^2(1-\eta^2)}\over 1-\eta^2},
\qquad \eta={t-y\over\sqrt{y^2+s^2}}$$
gives
$$\displaylines{\qquad
{d^l\over dt^l}I^{lm}(t)=8\pi(-1)^l\int_0^\infty{ds\over s}
\left(d\varphi\over ds\right)^2
\int_{t/2}^\infty dy{(s^2+yt)^2\over y^2 + s^2} \hfill\cr
\hfill\times\left[\sum_{i=-2,0,2} P_{l+i}\left
({t-y\over\sqrt{y^2+s^2}}\right)
\int d\Op A^{l+i,lm}(\Op) \Theta(\Op,y) \right].
\qquad(48)}$$
This expression can be simplified by noting that (1) $d\phi/ds$
is non-zero only when $s$ is smaller than the initial bubble
radius, and (2) the angular integral is zero until the bubble
first collides, which by assumption is only after the bubble has
expanded by a very large factor. Thus $y\gg s$ and to a very
good approximation $s$ can be set to zero in the integrand of
the $y$-integral. Now using Eq.~(9b), the $s$-integral can be
performed to give $\rv/3$; Eq.~(48) becomes
$${d^l\over dt^l}I^{lm}(t)={8\pi\over 3}(-1)^l\rv t^2
\int_{t/2}^\infty dy\sum_{i=-2,0,2} P_{l+i}\left({t\over y}
-1\right) \Theta^{l+i,lm}(y),\eqno(49)$$
where
$$\Theta^{l+i,lm}(t)\equiv\int d\Op A^{l+i,lm}(\Op)
\Theta(\Op,t).\eqno(50)$$
Note $\Theta^{l+i,lm}$ vanishes (by spherical symmetry) until
the bubble first collides; likewise, $\Theta^{l+i,lm}$ vanishes
for $t\geq R$, where $R$ is the size of the bubble when its
surface has completely collided, since $\Theta(\Omega,t\geq
R)=0$.
Analagous formulas holds for the ``current'' multipole
moments $S^{lm}$, but they vanish because the
tensor-spherical harmonics contracted with the unit vectors are
identically zero (see Appendix).

The multipole radiation from a single bubble is only
nonzero for $0<t<2R$.
This makes sense physically. The bubble wall propogates
outwards at essentially the speed of light;
even though the bubble does not begin to radiate
until it first collides, the first radiation
from the bubble still reaches an observer at distance
$r$ at time $t=r$. Likewise, radiation
generated on the opposite side of the bubble from the observer
will arrive at time $t=r + 2R$, since the diameter
of the bubble when it disappears is $2R$.

Equation (49) provides the key to computing the gravitational
radiation from a single bubble. The evaluation of
$d^lI^{lm}/dt^l$ only involves computing $\Theta^{l+i,lm}(t)$,
which depends upon the ``collision history'' of a given bubble.
We present two different estimates for $\Theta^{l+i,lm}(t)$.
The first is analytical,
based upon the fraction of the bubble surface that remains
uncollided at time $t$; the second estimate is derived from our
numerical simulations of bubble collisions.

Once we have $d^lI^{lm}/dt^l$ in hand for a single bubble,
it is a simple matter to
calculate the power and the spectrum of the
gravitational radiation by summing incoherently over the
distribution of bubbles:
$$\eqalignno{P(t)&={G\over 32\pi}\int_0^\infty
\sum_{l,m}\left|{d^{l+1}\over dt^{l+1}}I^{lm}(t,R)
\right|^2{dn\over dR} dR;&(51)\cr
{dE_{GW}\over d\omega}&={G\omega^2\over 8}
\int_0^\infty\sum_{l,m}
\left|{}^{(l)}I^{lm}(\omega,R)\right|^2
{dn\over dR} dR; &(52)}$$
where ${}^{(l)}I^{lm}(\omega,R)$ is the Fourier transform of
$d^lI^{lm}/dt^l(t,R)$. The $R$ in the argument refers to the
multipole radiation from a single bubble whose size at the end
of the phase transition is $R$, and $dn/dR$ is the distribution
of bubble sizes, {\it cf.}\ Eq.~(33).
Note that the
expressions in Eqs.~(51) and (52) are for power and energy per
unit volume (since $dn/dR$ is the bubble size distribution per
unit volume).

\medskip
\centerline{\sl (b) Analytic approximation to
$\Theta^{l+i,lm}(t)$}
\nobreak
\medskip

First we consider a very simple analytic approximation. Recall
that $\Theta^{l+i,lm}$ is an integral over the uncollided bubble
envelope. It must depend upon the fraction of the bubble wall
that remains uncollided at time $t$. This fraction, $f(t,t_R)$,
is given by [7]
$$f(t,t_R)=e^{-I(t)+I(t_R)}\eqno(53)$$
where $t_R$ is the nucleation time for a bubble which has
radius $R$ at the end of the phase transition,
{\it i.e.}, $t_R=t_* -R$. Since $\Theta^{l+i,lm}(t)$
must vanish at early times when the bubble has not yet collided
($f=1$), and at late times when the bubble surface is completely
collided ($f=0$), $\Theta^{l+i,lm}(t)$ can be expressed as a sum
of the terms $(1-f)^mf^n$ ($n,\,m=1,\,2,$\dots).
We take as a simple {\it ansatz}
$$\Theta^{l+i,lm}(t)=cf(1-f)\eqno(54)$$
where $c$ is an undetermined normalization constant.

We now estimate gravitational radiation using the quadrupole ($l=2$)
term of the multipole expansion. Taking only the simplest
($i=-2$) term of Eq.~(49), we approximate ${\ddot I}^{2m}(t)$ as
$${d^2\over dt^2}I^{2m}(t,R)={8\pi\over 3}c\rv u^2\int_{u/2}^\infty
f(u^\prime,R)\left[1-f(u^\prime,R)\right]du^\prime\eqno(55)$$
where we set $t_*=0$ as the time origin so that
$t_R=-R$, $u=t-t_R=t+R$, and $\Gamma\propto e^{\beta t}$, giving
$$f(u,R)=\exp\left[-Me^{\beta(u-R)}+Me^{-\beta
R}\right];\eqno(56)$$
recall $M=I(t_*)$. Note by setting $t_*=0$ instead of $t_R=0$,
and by using $u=t-t_R$, we take into account that bubbles of
different sizes are nucleated at different times.
The power radiated in gravitational waves per unit volume
is given by the
incoherent sum over the distribution of bubble sizes,
Eq.~(51); we have replaced the
sum over $m$ in Eq.~(51) with a factor of 5.
Likewise, the energy spectrum of gravitational waves per unit
volume is given by Eq.~(52).

In Fig.~9 we show $P(t)$ and in
Fig.~10 we show $\omega dE_{GW}/d\omega$. At low
frequencies, the spectrum behaves just as in our many-bubble
simulation; however, at high frequencies it falls off more
rapidly, and the peak of the spectrum is about a factor of three
lower. The deficiency in high frequency power traces to the fact
that we have neglected the sharp cusps which form in bubble
collisions. In this approximation the fraction of vacuum energy
released in gravity waves is $E_{GW}/E_{\rm
vac}=1.5c^2(H/\beta)^2$, so that $c\simeq 0.2$ reproduces the
result of our many-bubble simulations. We have tried a range
values for powers of $f$ and $1-f$ in Eq.~(54); these
alternatives change only the overall normalization of the
spectrum (decreasing with increasing powers).

\medskip
\centerline{\sl (c) Numerical estimation of
$\Theta^{l+i,lm}(t)$}
\nobreak
\medskip

It is straightforward to extract from our numerical simulations
average multipole moments for a single bubble,
and thereby determine a normalized approximate spectrum.
Specifically, for 70 individual bubbles,
nucleated in two of our five smaller simulations, we have
computed $d^lI^{lm}(t)/dt^l$ by using Eq.~(49), for $l=2$ and
$l=3$ (the quadrupole and octupole moments). These multipole
moments give the gravitational waveforms from a given bubble via
Eq.~(40); quadrupole and octupole waveforms for a representative
bubble are displayed in Fig.~11.
The radiated power $P(t)$ for the same bubble is
shown in Fig.~12; note that the octupole power is only around
10\% of the quadrupole power, so we can safely assume that
contributions from higher moments are negligible.

To calculate an ``average'' energy spectrum from the 70 bubbles, we
first construct scaled multipole moments, removing a factor of
$R$ from the time variable and a factor of $\rv R^3$ from
$d^lI^{lm}(t)/dt^l$ for each bubble. Then we calculate the
energy spectrum for each bubble using these scaled moments, and
average over the 70 bubbles to give the average spectrum for a
bubble. This average energy spectrum, shown in Fig.~13,
is quantitatively similar to the two-bubble
spectrum, peaking at around $\omega R=4.6$ with roughly the same
overall normalization, but dropping off much faster on the
high-frequency side of the peak. We then follow
the same procedure as in the previous subsection, integrating
over the bubble-size distribution to give the energy
spectrum per unit volume of the radiation from the phase transition.
The energy per octave is compared with the
spectrum from our many-bubble simulation in Fig.~10.

Overall, this statistical approximation does a reasonable job.
As with the previous analytical approximations, it
closely reproduces the low-frequency behavior.
The peak of the approximate spectrum has about the
correct amplitude, though the peak frequency is low
by about a factor of two.
The most obvious discrepancy is again the rapid high-frequency
drop of the approximate spectrum, which falls off as $\omega^{-6}$
for large frequencies, in marked contrast to the many-bubble
calculation, which falls off as $\omega^{-2.8}$. As before, the
deficiency of high-frequency power is due to the neglect of
cusps. In this approximation the fraction of energy radiated in
gravitational waves is
$E_{GW}/E_{\rm vac}=0.036(H/\beta)^2$, compared with
$0.06(H/\beta)^2$ in the many-bubble simulations.

The utility of our pair of statistical approximations lies in
their computational ease.
The difference in computing time between these
approximations and the many-bubble simulations is enormous.
In Fig.~10, the analytical statistical approximation required
negligible computing time, the numerical statistical approximation
used around an hour of computing time, while the many-bubble
points required several weeks on the same machine.
Furthermore, our averages for the radiation from a single bubble can
be used with any form for the bubble nucleation rate, by
substituting the appropriate form for $dn/dR$ in Eq.~(52). These
approximations give
a rough, but quick, estimate of the gravitational radiation
from any strongly first-order phase transition for which the
bubble nucleation rate is known.

\bigskip
\noindent{\bf V. Discussion and Concluding Remarks}
\nobreak
\bigskip

Before summarizing the present work, let us
place it in context by reviewing our previous work.
Based largely on dimensional estimates,
it was argued that the gravitational
radiation produced by the collision of vacuum bubbles
in a strongly first-order phase transition could
account for a substantial fraction of the vacuum energy
released [5]. This conjecture was verified
in our previous numerical work
[6] where we calculated the gravitational
radiation resulting from the collision of two vacuum bubbles
by evolving the scalar-field configuration corresponding
to two vacuum bubbles nucleated simultaneously and
separated by distance $d$.  This calculation
was carried out in the linearized-gravity
approximation and the expansion of
the universe was neglected, both assumptions being valid
provided that the duration of the transition is
less than a Hubble time.  We found that the amount of radiation
emitted is indeed significant
and only depends upon the duration of the
collision---and not the fine-scale details of the bubbles.
The fraction of vacuum energy liberated into gravitational waves
is $E_{\rm GW}/E_{\rm vac} \simeq
1.3 \times 10^{-3}\,(H\tau )^2$, valid for $\tau \sim d$,
where $\tau$ is the duration of the transition.
Unfortunately, this work depended upon the
phenomenological parameter $\tau$;  moreover, it is
a bold extrapolation to use the collision
of two bubbles to model a realistic
phase transition, which consists of many bubbles
of different sizes colliding. As noted earlier, it is beyond present
computational capabilities to collide more than
a few bubbles by scalar-field evolution.

These drawbacks led to the present work:  the development
of a workable approximation to study the gravitational radiation
from hundreds of colliding bubbles.  Motivated by
the fact that our two-bubble results only depend
upon the gross features of the collision, we developed
the envelope approximation described in this paper.  In the
envelope approximation an expanding bubble is
treated as a very thin shell of energy (equal to the
vacuum energy it liberates); when
bubbles collide only their envelope is followed
and their overlap (interaction) regions are ignored.  By considering
the collision of two bubbles we showed that the
envelope approximation accurately reproduces
our previous two-bubble results; {\it e.g.}, the
energy spectrum in gravitational waves agrees to around 20\%.

Having established the validity of the envelope
approximation, we nucleated hundreds of vacuum bubbles in
spherical volumes with a nucleation rate
that grows as $e^{\beta t}$:  specifically, 127 bubbles
total in five small simulations and 180 bubbles in
one large simulation.  (It is argued in Ref.~[7]
that such a functional
dependence for the bubble-nucleation rate applies with
great generality.)  Using the envelope approximation
we computed the fraction of vacuum energy released
in gravitational waves and found:  $E_{\rm GW}/E_{\rm vac}
\simeq 0.06\,(H/\beta )^2$ with the energy spectrum
peaking at a frequency $\omega \simeq 1.6\beta$ ($H^2=
8\pi G\rho_{\rm vac}/3$).  With this nucleation
rate the duration of the phase transition $\tau
\simeq 3/\beta$; using this fact, it follows that
the fraction of energy liberated in gravitational
waves is about five times the estimate based upon
our previous two-bubble results, with the spectrum
peaking at about the same frequency.
We believe that additional energy is released in
gravity waves because a given bubble
collides with many other bubbles
rather than a single bubble.

In the present work we also have developed two statistical
approximations that allow simple analytical or
semi-analytical approximations to the energy spectrum
radiated in gravitational waves.  Both approximations
provide better than order-of-magnitude accuracy and greater
ease of calculation, and are particularly well suited
to computing the gravitational radiation for an arbitrary
bubble nucleation rate.

On very general grounds it has been argued that
$\beta^{-1}$, which controls the time/length scale
of the phase transition, is of the order of a few
percent of $H^{-1}$ or greater (see Section IV or Ref.~[7]),
indicating that the fraction of vacuum energy liberated in gravitational waves
in a phase transition that proceeds through the nucleation and
collision of vacuum bubbles is of order $10^{-4}$ or so.
We have addressed the potential
observational consequences of our results in a {\it
Letter} [8]; very briefly, in terms of the temperature
of the universe after the phase transition,
the fraction of critical density contributed by
gravitational waves produced is $\Omega_{\rm GW} \sim
10^{-9}$ with characteristic frequency $f\sim
10^{-6}\,(T/\GeV )\,$Hz.  There we also discuss
the prospects for the detection of such a stochastic
background of gravitational waves with the coming
generation of laser interferometer gravity-wave
observatories [14].

Two key assumptions underlay all of our work:
(1) the use of linearized gravity and the neglect of the expansion
of the universe; and (2) the assumption that
the bubbles expand at constant
acceleration (put another way, all the vacuum energy liberated is
converted into the kinetic energy of the
bubble wall).  As we discussed in Ref.~[6] the first
assumption is valid so long as the duration of
the phase transition $\tau \sim \beta^{-1}$ is
much less than the Hubble time $H^{-1}$.
This should be satisfied for most phase transitions
as $\beta^{-1}$ is expected to be only a few percent
of $H^{-1}$.  However, there are situations where
this condition may not be satisfied, {\it e.g.}, in
some models of extended inflation [15]; moreover,
such situations are very interesting since our results
indicate that the fraction of energy liberated
in gravity waves approaches unity.  We are currently
trying to generalize our results by relaxing
the first assumption [16].

The second assumption is that all the vacuum energy
liberated by a bubble goes into the kinetic energy of its wall.
This is certainly true for a bubble nucleated at zero
temperature, but it may not be a good approximation
to one nucleated at finite temperature because of the
interaction of the bubble with the ambient thermal plasma.
The second assumption is certainly well justified
for models of first-order inflation where the universe has undergone
extreme supercooling during the inflationary epoch
so that the temperature of the Universe when the phase
transition occurs is exponentially small.  Whether or not
this assumption applies in a first-order phase
transition that only undergoes moderate supercooling
is an open question.  In this case it is
not implausible that much or even most of the
latent heat released is dissipated into heat
rather than the bulk motion of the expanding bubble
front (here we have used the term latent heat rather than vacuum energy).
The motion of a bubble wall in this circumstance
is not a simple matter to analyze:  both the microscopic
interaction of the ambient medium with the bubble
front and bulk hydrodynamics are important.
Though much work has been done the results are not conclusive [17].

The strength of a first-order phase transition can be
characterized by the ratio of the latent heat (per unit
volume) released to the energy density of the ambient
plasma, given by the fourth power of the temperature at which bubble
nucleation commences:  $\gamma=\rho_{\rm vac}/T_{\rm nuc}^4$; note that
the increase in entropy per comoving is proportional
to $\gamma^{3/4}$.  For an inflationary transition
$\gamma\rightarrow \infty$,
while for a weakly first-order transition $\gamma$ is of order unity or less.
For very large $\gamma$ it seems clear that the bubbles must behave
as vacuum bubbles (all the latent heat liberated goes into
accelerating the bubble wall).  What happens for moderate values of
$\gamma$ is still unclear and probably depends on the specific
phase transition under consideration. The latent heat could simply be
dissipated viscously, in which case little gravitational
radiation would be produced; or the latent heat
could be converted into the
bulk motion of the fluid (at some velocity less than
the speed of light), in
which case an appreciable amount of gravitational radiation
could still be produced.  This important issue is currently
under study [18].

\bigskip\bigskip
\centerline{\bf Acknowledgements}
\nobreak
\medskip

We thank Erick Weinberg, Rick Davis, John Preskill, and Paul
Steinhardt for helpful discussions.
This work has been supported by a National Science Foundation
graduate fellowship, the NASA Graduate Student Researchers
Program, by the DOE (at Chicago and Fermilab), and by NASA
through grant NAGW-2381 (at Fermilab).

\bigskip\bigskip
\centerline{\bf Appendix: Tensor Spherical Harmonics}
\nobreak
\medskip

In this appendix we review the formalism of
tensor-spherical harmonics,
and calculate the tensor contractions needed in
Sec. IV. Generally, we follow the notation
presented in [13]. A set of basis vectors $\xi^m$ are
coupled to form the traceless and symmetric basis
tensors ${\bf t}^m$:
$${\bf t}^m = \sum_{m^\prime = -1}^{+1} \sum_{m^{\prime\prime} =
-1}^{+1} (1\,1\,m^\prime\,m^{\prime\prime}|\,2\,m) \,\xi^{m^\prime}
\otimes \xi^{m^{\prime\prime}}\eqno(A.1a)$$
where $(l_1\,l_2\,m_1\,m_2|l_3\,m_3)$ is the Clebsch-Gordan
coefficient for adding angular momenta $l_1$ and $l_2$ to obtain $l_3$.
In terms of the Cartesian basis vectors ${\bf e}_x$,
${\bf e}_y$, and ${\bf e}_z$, these symmetric basis tensors are
$${\bf t}^{\pm 2}={1\over 2}({\bf e}_x\otimes {\bf e}_x
-{\bf e}_y\otimes {\bf e}_y)
\pm {i\over 2}({\bf e}_x\otimes {\bf e}_y +
{\bf e}_y\otimes {\bf e}_x);\eqno(A.1b)$$
$${\bf t}^{\pm 1}=\mp {1\over 2}({\bf e}_x\otimes
{\bf e}_z + {\bf e}_z\otimes {\bf e}_x)
- {i\over 2}({\bf e}_y\otimes {\bf e}_z +
{\bf e}_z\otimes {\bf e}_y);\eqno(A.1c)$$
$${\bf t}^0 = {1\over\sqrt 6}
(2{\bf e}_z\otimes {\bf e}_z - {\bf e}_x\otimes {\bf e}_x
-{\bf e}_y\otimes {\bf e}_y).\eqno(A.1d)$$
Then the relevant tensor spherical harmonics are
$${\rm T}^{2\,l^\prime,lm}=\sum_{m^\prime =-l^\prime}^{l^\prime}
\sum_{m^{\prime\prime}=-2}^2
(l^\prime\,2\,m^\prime\,m^{\prime\prime}|\,l\,m)Y^{l^\prime
m^\prime} {\bf t}^{m^{\prime\prime}},\eqno(A.2)$$
where $l^\prime =l,\,\,l\pm 1,\,\,l\pm 2.$
This represents the combination of an orbital angular momentum
$l^\prime$ and a spin angular momentum 2 to give total angular
momentum $l$. These spherical harmonics are eigenfunctions of
the orbital angular momentum operator ${\bf L}^2$ with eigenvalue
$l(l+1)$, like the more familiar $Y^{lm}$. Also, we have the
``pure spin tensor harmonics'' for spin 2,
$${\rm T}^{E2,lm}=a_2(l){\rm T}^{2\,l+2,lm}
+ a_0(l){\rm T}^{2\,l,lm}
+ a_{-2}(l){\rm T}^{2\,l-2,lm};\eqno(A.3a)$$
$${\rm T}^{B2,lm}=-ia_1(l){\rm T}^{2\,l+1,lm}
-ia_{-1}(l){\rm T}^{2\,l-1,lm};\eqno(A.3b)$$
where
$$\eqalignno{a_2(l)&\equiv\sqrt{l(l-1)\over 2(2l+1)(2l+3)}\,\,;&(A.4a)\cr
a_1(l)&\equiv\sqrt{l-1\over 2l+1}\,\,;&(A.4b)\cr
a_0(l)&\equiv\sqrt{3(l-1)(l+2)\over (2l-1)(2l+3)}\,\,;&(A.4c)\cr
a_{-1}(l)&\equiv\sqrt{l+2\over 2l+1}\,\,;&(A.4d)\cr
a_{-2}(l)&\equiv\sqrt{(l+1)(l+2)\over 2(2l+1)(2l-1)}\,\,.&(A.4e)\cr}$$
Under rotations around the radial vector, the harmonics (A.3)
transform like
the components of the polarization tensor of a pure spin-2
state, but they are not orbital angular momentum eigenfunctions.

Using the above definitions, it is straightforward to calculate
the explicit forms for the necessary tensor
contractions.
Equation (47) defines the auxiliary quantity
$$A^{l^\prime,lm}(\Omega)\equiv a_{l^\prime -l}(l)
{\rm T}^{2\,l^\prime,lm}_{pq}(\Omega)^*
\xhat_p(\Omega)\xhat_q(\Omega)$$
where $\xhat$ is a unit vector in the direction $\Omega$.
The expressions for $m<0$ follow from the identity
$${\rm T}^{2\,l^\prime,lm}=(-1)^{l^\prime +l+m}
({\rm T}^{2\,l^\prime,l\,-m})^*\eqno(A.5)$$

For $l=2$ (quadrupole moments):
$$A^{l^\prime,2m}=0\qquad{\rm for}\qquad l^\prime=1,
\quad l^\prime=3\eqno(A.6)$$
$$A^{0,22}={1\over 4}\sqrt{2\over 5\pi}
e^{-2i\phi}\sin^2\theta\eqno(A.7)$$
$$A^{2,22}=-{5\over 28}\sqrt{2\over 5\pi}
e^{-2i\phi}(1+\sin^2\theta+2\sin^4\theta)\eqno(A.8)$$
$$A^{4,22}={3\over 28}\sqrt{2\over 5\pi}
e^{-2i\phi}\sin^2\theta\eqno(A.9)$$
\smallskip
$$A^{0,21}=-{1\over 4}\sqrt{2\over 5\pi}
e^{-i\phi}\sin 2\theta\eqno(A.10)$$
$$A^{2,21}={5\over 112}\sqrt{2\over 5\pi}
e^{-i\phi}(8-2\cos 2\theta -\sin 4\theta)\eqno(A.11)$$
$$A^{4,21}=-{3\over 28}\sqrt{2\over 5\pi}
e^{-i\phi}\sin 2\theta\eqno(A.12)$$
\smallskip
$$A^{0,20}=-{1\over 6}\sqrt{3\over
5\pi}(1-3\cos^2\theta)\eqno(A.13)$$
$$A^{2,20}=-{5\over 42}\sqrt{3\over
5\pi}(1+3\cos 2\theta)\eqno(A.14)$$
$$A^{4,20}={1\over 28}\sqrt{3\over 5\pi}
(1+3\cos 2\theta)\eqno(A.15)$$
\bigskip
For $l=3$ (octupole moments):
$$A^{l^\prime,3m}=0\qquad{\rm for}\qquad l^\prime=2,
\quad l^\prime=4\eqno(A.16)$$
$$A^{1,33}=-{1\over 4}\sqrt{3\over 7\pi}
e^{-3i\phi}\sin^3\theta\eqno(A.17)$$
$$A^{3,33}={7\over 18}\sqrt{3\over 7\pi}
e^{-3i\phi}\sin^3\theta\eqno(A.18)$$
$$A^{5,33}=-{5\over 36}\sqrt{3\over 7\pi}
e^{-3i\phi}\sin^3\theta\eqno(A.19)$$
\smallskip
$$A^{1,32}={3\over 4}\sqrt{2\over 7\pi}
e^{-2i\phi}\cos\theta\sin^2\theta\eqno(A.20)$$
$$A^{3,32}=-{7\over 6}\sqrt{2\over 7\pi}
e^{-2i\phi}\cos\theta\sin^2\theta\eqno(A.21)$$
$$A^{5,32}={5\over 12}\sqrt{2\over 7\pi}
e^{-2i\phi}\cos\theta\sin^2\theta\eqno(A.22)$$
\smallskip
$$A^{1,31}=-\sqrt{5\over 7\pi}e^{-i\phi}\sin\theta
\left[{9\over 40} + {3\over 8}\cos 2\theta\right]\eqno(A.23)$$
$$A^{3,31}=-\sqrt{5\over 7\pi}e^{-i\phi}\sin\theta
\left[{21\over 80} + {7\over 12}\cos 2\theta
- {7\over 96}\cos 4\theta\right]\eqno(A.24)$$
$$A^{5,31}=-\sqrt{5\over 7\pi}e^{-i\phi}\sin\theta
\left[{1\over 8} + {5\over 24}\cos 2\theta\right]\eqno(A.25)$$
\smallskip
$$A^{1,30}=\sqrt{5\over 21\pi}\cos\theta\left[{3\over 4}\cos
2\theta - {3\over 20}\right]\eqno(A.26)$$
$$A^{3,30}=-\sqrt{5\over 21\pi}\left[{7\over 20}\cos\theta
+ {7\over 12}\cos 3\theta\right]\eqno(A.27)$$
$$A^{5,30}=\sqrt{5\over 21\pi}\left[{1\over 8}\cos\theta
+ {5\over 24}\cos 3\theta\right]\eqno(A.28)$$

\bigskip\bigskip
\centerline{\bf References}
\nobreak
\medskip

\item{[1]} V. A. Rubakov, M. Sazhin, and A. Veryaskin, Phys.
Lett. {\bf 115B}, 189 (1982); R. Fabbri and M. Pollack, {\it
ibid.} {\bf 125B}, 445 (1983).

\item{[2]} T. Vachaspati and A. Vilenkin, Phys. Rev. D {\bf 31},
3052 (1985); T. Vachaspati, A. E. Everett, and A. Vilenkin, {\it
ibid.} {\bf 30}, 2046 (1984); B. Allen and E. P. S. Shellard,
{\it ibid.} {\bf 45}, 1898 (1992); R. R. Caldwell and B. Allen,
{\it ibid.} {\bf 45}, 3447 (1992).

\item{[3]} B. J. Carr, Astron. and Astrophys. {\bf 89}, 6
(1980).

\item{[4]} C. J. Hogan, Mon. Not. R. Astron. Soc. {\bf 218}, 629
(1986); E. Witten, Phys. Rev. D {\bf 30}, 272 (1984); L. M.
Krauss, Phys. Lett. {\bf B284}, 229 (1992).

\item{[5]} M. S. Turner and F. Wilczek, Phys. Rev. Lett. {\bf
65}, 3080 (1990).

\item{[6]} A.~Kosowsky, M.~S.~Turner, and R.~Watkins, Phys. Rev.
D {\bf 45}, 4514 (1992).

\item{[7]} M. S. Turner, E. J. Weinberg, and L. M. Widrow,
Phys. Rev. D {\bf 46}, 2384 (1992).

\item{[8]} A. Kosowsky, M.S. Turner, and R. Watkins, Phys. Rev.
Lett. {\bf 69}, 2018 (1992).

\item{[9]} S. Coleman, Phys. Rev. D {\bf 15}, 2929 (1977); C.
Callan and S. Coleman, {\it ibid.} {\bf 16}, 1762 (1977).

\item{[10]} For a detailed discussion of vacuum bubble
kinematics, see S.~W.~Hawking, J.~M.~Stewart, and I.~G.~Moss,
Phys. Rev. D {\bf 26}, 2681 (1982); also, R. Watkins and L. Widrow, Nuc.
Phys. B {\bf 374}, 446 (1992).

\item{[11]} S. Weinberg, {\sl Gravitation and Cosmology} (Wiley,
New York, 1972), Ch.~10.

\item{[12]} A. H. Guth and H. Tye, Phys. Rev. Lett. {\bf 44},
631 (1980); {\it ibid.} {\bf 44}, 963(E) (1980).

\item{[13]} K. S. Thorne, Rev. Mod. Phys. {\bf 52}, 299 (1980).

\item{[14]} A.~Abramovici {\it et al.}, {\it Science} {\bf 256},
325 (1992); J. E. Faller {\it et al.}, Adv. Space Res. {\bf 9},
107 (1989).

\item{[15]} For a discussion of extended inflation, see D. La and
P. J. Steinhardt, Phys. Rev. Lett. {\bf 62}, 376 (1989); E. J.
Weinberg, Phys. Rev. D {\bf 40}, 3950 (1989); and E. W. Kolb,
Phys. Scr. {\bf T36}, 199 (1991).

\item{[16]} A.~Kosowsky, M.S.~Turner, and E.~Weinberg,
unpublished.

\item{[17]} D. A. Kirzhnits, JETP Lett. {\bf 15}, 529 (1972); D.
A. Kirzhnits and A. D. Linde, Ann. Phys. {\bf 101}, 195 (1976);
M. Dine {\it et al.}, Phys. Rev. D {\bf 46}, 550 (1992);
M. Gyulassy {\it et al.}, Nuc. Phys B {\bf 237},
477 (1984); K. Enqvist {\it et al.}, Phys. Rev. D {\bf 45},
3415 (1992); M. Kamionkowski and K. Freese, Institute for
Advanced Studies Report No. IASSNS-HEP-92/46 (1992),
unpublished; P. J. Steinhardt, Phys. Rev. D {\bf 25}, 2082
(1982); B.~Liu, L.~McLerran, and N.~Turok, Phys. Rev. D {\bf
46}, 2668 (1992).

\item{[18]} M.~Kamionkowski, A.~Kosowsky, and M.S.~Turner,
unpublished.

\vfill
\eject

\centerline{\bf Figure Captions}
\bigskip

{\bf Figure 1}: The kinetic (dashed curve), gradient (dot-dashed
curve), and total (solid curve) energy of a
single expanding vacuum bubble. The scales are
arbitrary; the bubble has initial radius of about 10, in the
same units as $t$. Note that by $t=20$, when the bubble has approximately
doubled in size, the total energy scales almost exactly as
$t^3$, the vacuum energy liberated by a bubble expanding at the
speed of light from zero initial size, and resides equally in
the kinetic and gradient energies of the bubble wall.

\medskip
{\bf Figure 2}: A schematic picture that illustrates the
envelope approximation. The dark lines are the bubble walls,
expanding at the speed of light. The shaded areas are the
interaction regions; the envelope approximation
neglects the interaction regions and takes into account only the
bubble envelopes. Snapshot (b) is at a somewhat later time, and
three new bubbles have been nucleated.

\medskip
{\bf Figure 3}: The energy spectrum of radiation from two
colliding bubbles, in the quadrupole approximation.
The units are the same for both curves, but arbitrary.
The dashed curve is the
result from detailed scalar field evolution (Ref. [6]), and the
solid curve from the envelope approximation.

\medskip
{\bf Figure 4}: The energy spectrum from two colliding bubbles
in the full linearized-gravity approximation. The units are
the same as in Fig.~3.
The solid line is the envelope approximation,
which reproduces well the results from detailed
scalar-field evolution [6], the dashed curve.

\medskip
{\bf Figure 5}: The distribution of bubble sizes,
both unweighted and weighted by the bubble's total energy, for
the exponential nucleation rate $\Gamma(t)=\Gamma_0e^{\beta t}$.
The energy-weighted distribution peaks at a radius that is
almost twice as large.

\medskip
{\bf Figure 6}: A slice through a spherical sample volume at
different times.
The volume boundary is the outer circle;
the bubble walls are the darker
curves within the boundary. A
total of 33 bubbles were nucleated in this volume during the
phase transition.

\medskip
{\bf Figure 7}: The differential energy spectrum $dE/d\omega
d\Omega$, for three orthogonal directions in a simulation with
33 colliding bubbles. Note the large variations in the
different directions.

\medskip
{\bf Figure 8}: The energy spectrum per
frequency octave, divided by the vacuum energy of the sample
volume. The points with error flags are averaged over the five
different simulations and integrated over six directions
per simulation. The error bars reflect the standard estimate for
the deviation of
the mean. The dashed line is the spectrum for two bubbles,
calculated using scalar-field evolution [6], with
$\tau=3.2\beta$. The solid triangles are the results of the
180-bubble simulation (integrated over six directions).

\medskip
{\bf Figure 9}: The radiated power $P(t)$ per unit volume
in the first statistical
approximation. The phase transition completes at $t=0$.

\medskip
{\bf Figure 10}: The energy spectrum as computed in the first
statistical approximation, with $c=1$ (solid line) and the
spectrum from the second statistical approximation (dashed line),
compared with the results
of our many-bubble simulations (points with error flags).

\medskip
{\bf Figure 11}: The quadrupole and octupole moments for a given
bubble nucleated at time zero. $R$ is the final bubble radius.
(a) The real and imaginary parts of the quadrupole moments for
$m=0,\, 1,\, 2$ (the imaginary part of the $m=0$ moments
vanish). (b) The same for the octupole moments for $m=0,\, 1,\,
2,\, 3$.

\medskip
{\bf Figure 12}: The quadrupole (upper curve) and octupole
(lower curve) contributions to the total power radiated from the
bubble in Fig.~10.

\medskip
{\bf Figure 13}: The mean energy spectrum of a ``typical''
bubble, derived by averaging over 70 individual bubbles.
The standard estimate for the deviation of the mean is around
$\pm 10$\%.

\end